\documentclass[11pt]{article}
\usepackage{amssymb}
\usepackage{amsmath}
\usepackage{graphicx}
\usepackage{bm}
\newcommand{\mathsym}[1]{{}}

\usepackage{graphicx}
\usepackage{rotating}
\usepackage{color}
\usepackage{cancel}

\newcommand{\bra}{\begin{array}}
\newcommand{\era}{\end{array}}
\newcommand{\beq}{\begin{equation}}
\newcommand{\eeq}{\end{equation}}
\newcommand{\beqar}{\begin{eqnarray}}
\newcommand{\eeqar}{\end{eqnarray}}

\def\BC{\bb C}
\def\_\BC{\bbi C}

\def\( {\left(}
   \def\) {\right)}
\def\[ {\left[}
\def\] {\right]}

\def\dag {{\dagger}}

\begin{document}

\vspace{20pt}

\begin{center}

{\LARGE \bf New  Deformation of  quantum oscillator algebra: 
Representation and some applications\\
\medskip
 }
\vspace{15pt}

{\large S. Arjika$^{\dag,1}$,  D. Ousmane Samary$^{\ddag,1,2}$,\\ E. Balo\"itcha$^{*,1}$ and M. N. Hounkonnou$^{**,1}$ } 

\vspace{15pt}

{\sl $1$-International Chair in Mathematical Physics and Applications\\ (ICMPA-UNESCO Chair), University of Abomey-Calavi,\\
072B.P.50, Cotonou, Rep. of Benin}\\

{\sl $2$-CNRS - Universit\'e Lyon 1, Institut Camille Jordan\\ (Bat. Jean Braconnier,  bd du 11 novembre 1918),\\
F-69622 Villeurbanne Cedex, France}\\

\vspace{5pt}
E-mails:  { $^{\dag}$rjksama2008@gmail.com, \sl 
$^\ddag$ousmanesamarydine@yahoo.fr, \\
\sl $^*$ezinvi.baloitcha@cipma.uac.bj, \sl 
$^{**}$norbert.hounkonnou@cipma.uac.bj}

\vspace{10pt}
\end{center}

%
%
%
%
%
%
%
%
%
%

\date{\today}
\begin{center}
[Ref-preprint]
CIPMA-MPA/031/2012
\end{center}
\begin{abstract}
This work addresses the study of the  oscillator algebra, defined by four  parameters $p$, $q$, $\alpha$, and $\nu$.  The time-independent Schr\"{o}dinger equation for the induced deformed harmonic oscillator is  solved;  explicit
analytic expressions of  
the energy spectrum   are given.   Deformed states are built and discussed
with respect to the  criteria of coherent state construction. 
 Various commutators involving annihilation and creation operators and their combinatorics are computed 
and analyzed.  Finally,
 the correlation functions of matrix elements of main normal and antinormal forms, pertinent for  quantum optics
 analysis,  are computed.
\end{abstract}


\section{Introduction}
\label{sect:intro}
The deformation of the harmonic oscillator algebra  whose applications in physics are presently 
rather technical but nonetheless very promising, possesses an important and useful representation theory
  in  connection to that of their classical limit algebra. From the other side, there are some hopes that,
 in physical studies of nonlinear phenomena, the deformed oscillator can play the same role   as the usual boson oscillator in
 usual nonrelativistic quantum mechanics. This could explain why various quantum deformations of boson oscillator commutation relations
  have attracted a great attention during the last few years (see 
\cite{Balo}-
\cite{Wei}  and references therein).
 This might be also due to the fact that there exist correspondences between quantum groups, quantum algebras, 
statistical mechanics, quantum field theory, conformal field theory, quantum and nonlinear optics  and noncommutative geometry, etc. 
 Furthermore, such a connection is extended to
coherent states (cs) deducible from
the  study of quantum groups and, therefore, from 
 the deformation of Heisenberg algebra.  
 As a pertinent application in quantum optics, cs can be used
  to compute matrix elements, $_\nu\langle z|a^{\dag m} a^n|z\rangle_\nu,$ \,$_\nu\langle z|a^n a^{\dag m} |z\rangle_\nu,$
 corresponding to normal and antinormal forms, respectively  \cite{Wei},
 (also called {\it symmetric form}),   by using the  normal product technique.

 Note that the most spread, in the literature, multiparameter deformation of the 
 harmonic oscillator is the so-called $(p,q)-$deformed oscillator.
 For more details, see  \cite{BMK}. To cite an example, let us mention
the $(p,q)-$deformed oscillator algebra   defined by the following Chakrabarti and Jagannathan's commutation relations
 \cite{chak}:

$
aa^\dagger-q a^\dagger a= p^{-N}, \quad aa^\dagger-p^{-1} a^\dagger a= q^{N},
$
\,
$
[N,a]=-a,\quad [N,a^\dagger]=a^\dagger,
$\,
where $p,q\in\mathbb{R};$ $a, a^\dagger$ and $N=a^\dagger a$ are the annihilation, creation and number operators, respectively. 
This deformation has been  generalized in various ways in the literature \cite{Balo} 
\cite{imb} and \cite{Kinani}. 

 This paper extend the result on \cite{Kinani} by introducing two new positive parameters functions $\phi_1$  and $\phi_2$. The spectrum and states of this new class of quantum oscillator algebra are given. We also analyzed the coherent states and correlation functions useful for the study of quantum optics properties.

 The work is organized as follows. In section 2, we give the preliminary  and definitions permitted to define properly  the mean result of our work.   Section 3 is  devoted to 
 compute related deformed states.  Relevant correlation functions as well as new identities are also built in section 4.  We give in Section 5 a discussion of a particular case of algebra \eqref{sa7}. We end with the   concluding  and  remarks in section 6.

\section{New class of deformation quantum oscillator algebra} 
We start from the following   mains definitions of the $ (p, q)-$deformed
oscillator algebra and its generalization.

$\bullet$ 
The $(p,q)-$oscillator algebra is generated by three elements $a,\;a^\dag$ and $N$ obeying
the relation \cite{chak}
\begin{eqnarray}
\label{sa3}
aa^\dag-p^{-1}a^\dag a=q^N,\;\; aa^\dag-qa^\dag a=p^{-N},\;\;
[N, a] = - a,\;\;
[N, a^\dag ] =  a^\dag. 
\end{eqnarray}

$\bullet$ The   $(q,p;\alpha,\beta,l)-$deformed canonical commutation relations defined in \cite{imb} is written as follows
\begin{eqnarray}
\label{sa5}
a a^\dag - q^l a^\dag a =p^{-\alpha N -\beta},\; a a^\dag-
p^{-l} a^\dag a = q^{\alpha N +\beta}\cr
[N, a] = -\frac{l}{\alpha} a,\;
[N, a^\dag ] =\frac{l}{\alpha}  a^\dag.
\end{eqnarray}
It is worth noticing that further generalization involving a new parameters $\phi_1$ and $\phi_2$ generate a richer algebra,
 with novel interesting properties, as we will see in the sequel. In this case, we arrive at  the following definition.
\\
{\it {\bf Definition1: (The deformed  algebra) }
The 
deformed
 algebra  generated by the operators $\{1, a, a^\dag , N\}$  is defined in this paper by the relations
\begin{eqnarray}
\label{sa7}
a a^\dagger - q^\nu a^\dagger a =\phi_1(p,q)
p^{-\alpha N }, \quad
 a a^\dagger - p^{-\nu} a^\dagger a = \phi_2(p,q)
q^{\alpha N },
\end{eqnarray}
\begin{eqnarray}
\label{eqref}
[ N, a^\dagger ] = \frac{\nu}{\alpha} a^\dagger,\;\; \;[ N, a] = -
\frac{\nu}{\alpha} a.
\end{eqnarray}
where $\phi_1(p,q)$ and $\phi_2(p,q)$ are two non singular and real valued positive functions of deformation parameters $p$ and $q$.}
\\
{\it {\bf Remark1:} It's important to notify immediately that in the limit  when $\nu,\alpha\to 1$ and $\phi_1(p,q)= \phi_2(p,q)=1$, one recovers the algebra studied by Chakrabarti et al \cite{chak}.  In the same manner if $\phi_1(p,q)=p^{-\beta},\phi_2(p,q)=q^\beta$, we recover 
the well known  algebra investigated by Burban \cite{imb}.  So the introduction of such regular functions $\phi_1$ and $\phi_2$ generalized the  previous oscilator algebras existing in the literature.}

From \eqref{sa7}, one can easily deduce that
 \begin{eqnarray}
\label{sa8}
aa^\dagger =
\frac{\phi_1(p,q)p^{-\alpha N -\nu}-\phi_2(p,q)q^{\alpha N +\nu}}{p^{-\nu}-q^\nu},
\,
a^\dagger a= \frac{\phi_1(p,q)
p^{-\alpha N}-\phi_2(p,q)q^{\alpha N }}{p^{-\nu}-q^\nu}.
\end{eqnarray}
Let  $\mathcal{F}$ be a Fock space
spanned by the orthogonal basis $\{|n\rangle, n=0, 1, 2 \ldots\}$. 
We defined the vacuum state $|0\rangle$ as the eigen-vector of $N$ such that $N|0\rangle=\chi_0|0\rangle.$ Due to the commutativity
 of $aa^\dag$ and $a^\dag a$ with $N$, we may assume that $aa^\dag|0\rangle=\mu_0|0\rangle, \quad a^\dag a|0\rangle=\lambda_0|0\rangle$, $\chi_0, \,\mu_0$ and $\lambda_0$ are three reals numbers. We will also impose the normalization   condition $\langle 0|0\rangle=1.$ 
\\
{\it {\bf Proposition1 }
The states $|n\rangle$ are built as follows: if $p q <1$ and $\phi_2(p,q)<\phi_1(p,q)$
\begin{eqnarray}
\label{ket1}
|n\rangle=
\frac{p^{n\nu(n-1)/4}}{\sqrt{\tau_{p^{-1},q}^n(\phi_1)(\phi_1^{-1}(p,q)\phi_2(p,q)(p q)^{\alpha\chi_0+\nu};(pq)^\nu)_n}}a^{\dag n}|0\rangle,\quad 
n\geq 0,
\end{eqnarray}
and if $p q >1$ and $\phi_1(p,q)<\phi_2(p,q)$,
\begin{eqnarray}
\label{ket2}
|n\rangle=
\frac{q^{-n\nu(n-1)/4}}{\sqrt{\tau_{q,p^{-1}}^n(\phi_2)(\phi_2^{-1}(p,q)\phi_1(p,q)(p q)^{-\alpha\chi_0-\nu};(p q)^{-\nu})_n}}a^{\dag n}|0\rangle,\quad 
n\geq 0.
\end{eqnarray}
The states \eqref{ket1} and \eqref{ket2} 
 satisfy the orthogonality and completeness conditions
\begin{eqnarray}
 \langle m|n\rangle=\delta_{mn}, \quad \sum_{n=0}^\infty |n\rangle \langle n|= \mathbf{1}.
\end{eqnarray}}

In order to understand and study properly this new oscillator  algebra we calculate 
the actions of the deformed operators $a$, $a^\dagger$ and $N$ on  $\mathcal{F}$ and get
\begin{itemize}
\item if $pq <1$ and $\phi_2(p,q)<\phi_1(p,q)$
\begin{eqnarray}
\label{asct}
a|n\rangle=\tau_{p^{-1},q}^{1/2}(\phi_1(p,q))p^{-(n-1)\nu/2}\sqrt{1-\frac{\phi_2(p,q)}{\phi_1(p,q)}(pq)^{\alpha\chi_0+n\nu}}\,|n-1\rangle.
\end{eqnarray}
\item if $pq >1$ and $\phi_1(p,q)<\phi_2(p,q)$
\begin{eqnarray}
\label{asct1}
a|n\rangle=\tau_{q,p^{-1}}^{1/2}(\phi_2(p,q))q^{(n-1)\nu/2}\sqrt{1-\frac{\phi_1(p,q)}{\phi_2(p,q)}(pq)^{-\alpha\chi_0-n\nu}}\,|n-1\rangle.
\end{eqnarray}
\item if $pq <1$ and $\phi_2(p,q)<\phi_1(p,q)$
\begin{eqnarray}
\label{asct2}
a^\dag|n\rangle=\tau_{p^{-1},q}^{1/2}(\phi_1(p,q))p^{-n\nu/2}\sqrt{1-\frac{\phi_2(p,q)}{\phi_1(p,q)}(pq)^{\alpha\chi_0+(n+1)\nu}}\,|n+1\rangle.
\end{eqnarray}
\item if $pq >1$ and $\phi_1(p,q)<\phi_2(p,q)$
\begin{eqnarray}
\label{asct3}
a^\dag|n\rangle=\tau_{q,p^{-1}}^{1/2}(\phi_2(p,q))q^{n\nu/2}\sqrt{1-\frac{\phi_1(p,q)}{\phi_2(p,q)}(pq)^{-\alpha\chi_0-(n+1)\nu}}\,|n+1\rangle.
\end{eqnarray}
\end{itemize}
Let  $\displaystyle \tau_{p,q}(t)=\frac{tp^{\alpha\chi_0+\nu}}{p^{\nu}-q^\nu},$ it gets even
\begin{eqnarray}
\label{asct4}
N|n\rangle&=&(\chi_0+n)|n\rangle.
\end{eqnarray}
From \eqref{ket1}-\eqref{asct3}, one can deduce that
\begin{eqnarray}
\label{mu}
aa^\dagger |n\rangle&=&
\frac{\phi_1(p,q)p^{-\alpha \chi_0-(n+1)\nu}-\phi_2(p,q)q^{\alpha \chi_0 +(n+1)\nu}}{p^{-\nu}-q^\nu}|n\rangle,\\
\label{mu1}
a^\dagger a|n\rangle&=& \frac{\phi_1(p,q)
p^{-\alpha \chi_0-n\nu}-\phi_2(p,q)q^{\alpha \chi_0+n\nu}}{p^{-\nu}-q^\nu}|n\rangle.
\end{eqnarray}
Finally we may conclude that, the states $|n\rangle$ solve the time-independent  Schr\"odringer equation of the deformed oscillator 
Hamiltonian $H =
a^\dag a + a a^\dag,$  i.e.
$
\label{aj}
H|n\rangle=
E_{n,\alpha}^{\nu} (p, q)|n\rangle
$, with  the corresponding  eigenvalue  $E_{n,\alpha}^{\nu} (p, q)$ given by 
\begin{eqnarray}
\label{ea1}
E_{n,\alpha}^{\nu} (p, q)
&=&\tau_{p^{-1},q}(\phi_1(p,q))p^{-(n-1)\nu}\Bigg\{1+p^{-\nu}-\frac{\phi_2(p,q)}{\phi_1(p,q)}(pq)^{\alpha\chi_0+n\nu}(1+q^{\nu})\Bigg\},\\
&&\mbox{ with }\,  pq <1 \mbox{ and } \phi_2(p,q)<\phi_1(p,q)\nonumber
\end{eqnarray}
\begin{eqnarray}
\label{ea2}
E_{n,\alpha}^{\nu} (p, q)&=&
\tau_{q,p^{-1}}(\phi_2(p,q))q^{(n-1)\nu}\Bigg\{1+q^{\nu}-\frac{\phi_1(p,q)}{\phi_2(p,q)}(pq)^{\alpha\chi_0+n\nu}(1+p^{-\nu})\Bigg\},\\
&&\mbox{ with }\, pq >1  \mbox{ and } \phi_1(p,q)<\phi_2(p,q).
\end{eqnarray}
It follows from \eqref{ea1} and \eqref{ea2} that the spectrum of the deformed Hamiltonian $H$
is symmetric under the change $q\to p^{-1}, \phi_1(p,q)\to\phi_2(p,q).$
\\
{\it{\bf Proposition2}
 For $A$ and $B\in L(\mathcal{L})$, where $ L(\mathcal{L})$ is  the set of linears operators acting on Fock space $\mathcal{L}$ we  consider the $q^\nu$ and $p^{-\nu}$ commutators defined by: 
$[A,B]_{q^{\nu}}=AB-q^{\nu}BA$  and 
$ [A,B]_{p^{-\nu}}=AB-p^{-\nu}BA,$
the following  brackets hold
\begin{eqnarray}
\label{sa11}
&[a,a^{\dag m+1}]_{q^\nu}=a^{\dag m}\Bigg(
\frac{\phi_1(p,q)p^{-\alpha N}(p^{-(m+1)\nu}-q^\nu)-\phi_2(p,q)q^{\alpha N}
(q^{(m+1)\nu}-q^\nu)}{p^{-\nu}-q^\nu}\Bigg),\cr
\label{sa12}
&[a,a^{\dag m+1}]_{p^{-\nu}}=a^{\dag m}\Bigg(
\frac{\phi_1(p,q)p^{-\alpha N}(p^{-(m+1)\nu}-p^{-\nu})-\phi_2(p,q)q^{\alpha N}
(q^{(m+1)\nu}-p^{-\nu})}{p^{-\nu}-q^\nu}\Bigg).
\end{eqnarray}}
{\bf Proof} The proof can be performed by using equation \eqref{sa8}. $\blacksquare$

For
$n, m\in\mathbb{N}\diagdown\{0\}$,  the expressions of operators $a^n a^{\dag m}$ and $a^{\dag m} a^n$ are given in  different cases by the following relations

 $\bullet\,\,\,\,\,\,$For $n < m$
\begin{eqnarray}
\label{saa}
a^n a^{\dag m} 
=
p^{-\nu\binom{n}2}\mathcal{T}_{p^{-1},q}^n(\phi_1(p,q))\Big(\frac{\phi_2(p,q)}{\phi_1(p,q)}(p q)^{\alpha N+\nu};(pq)^{\nu}\Big)_n a^{\dag m-n}
\end{eqnarray}
if $pq <1$ and $\phi_2(p,q)<\phi_1(p,q)$.
\begin{eqnarray}
a^n a^{\dag m} 
=
q^{\nu\binom{n}2}\mathcal{T}_{q,p^{-1}}^n(\phi_2(p,q))\Big(\frac{\phi_1(p,q)}{\phi_2(p,q)}(p q)^{-\alpha N-\nu};(pq)^{-\nu}\Big)_n a^{\dag m-n},
\end{eqnarray}
if $pq >1$ and $\phi_1(p,q)<\phi_2(p,q)$.
\begin{eqnarray}
\label{saaa}
a^{\dag m} a^n=
p^{-\nu\binom{n}2}\Big[{_{(p,q)}}\mathcal{S}_{a^\dag}^{-m}\mathcal{T}_{p^{-1},q}(\phi_1(p,q))\Big]^n\Big(\frac{\phi_2(p,q)}{\phi_1(p,q)}(p q)^{\alpha N+\nu};(pq)^{\nu}\Big)_n a^{\dag m-n},
\end{eqnarray}
if $pq <1$ and $\phi_2(p,q)<\phi_1(p,q)$.
\begin{eqnarray}
\label{saasaa}
a^{\dag m} a^n=
q^{\nu\binom{n}2}{_{(p,q)}}\mathcal{S}_{a^\dag}^{-m}\mathcal{T}_{q,p^{-1}}^n(\phi_2(p,q))\Big(\frac{\phi_1(p,q)}{\phi_2(p,q)}(p q)^{-\alpha N-\nu};(pq)^{-\nu}\Big)_n a^{\dag m-n},
\end{eqnarray}
if $pq >1$ and $\phi_1(p,q)<\phi_2(p,q)$.

 $\bullet\,\,\,\,\,\,$If $n > m$,
\begin{eqnarray}
\label{saa1}
a^n a^{\dag m} = 
p^{\nu\binom{m}2}{_{(p,q)}}\mathcal{S}_{a^\dag}^n\Bigg(\Big[{_p}\mathcal{S}_{a^\dag}^{-1}\mathcal{T}_{p^{-1},q}^m((\phi_1(p,q))\Big]\Big(\frac{\phi_2(p,q)}{\phi_1(p,q)}(p q)^{\alpha N};(pq)^{-\nu}\Big)_m\Bigg)a^{n-m},
\end{eqnarray}
if $pq <1$ and $\phi_2(p,q)<\phi_1(p,q)$.
\begin{eqnarray}
a^n a^{\dag m}  = 
q^{-\nu\binom{m}2}{_{(p,q)}}\mathcal{S}_{a^\dag}^n\Bigg(\Big[{_q}\mathcal{S}_{a^\dag}^{-1}
\mathcal{T}_{q,p^{-1}}^m((\phi_2(p,q))\Big]\Big(\frac{\phi_1(p,q)}{\phi_2(p,q)}(p q)^{-\alpha N};
(pq)^{\nu}\Big)_m\Bigg)a^{n-m},
\end{eqnarray}
if $pq >1$ and $\phi_1(p,q)<\phi_2(p,q)$.
\begin{eqnarray}
\label{dine}
a^{\dag m} a^n
&=&p^{\nu\binom{m}2}\Big[{_p}\mathcal{S}_{a^\dag}^{-1}\mathcal{T}_{p^{-1},q}^m(\phi_1(p,q))\Big]\Big(\frac{\phi_2(p,q)}{\phi_1(p,q)}(p q)^{\alpha N};(pq)^{-\nu}\Big)_m a^{n-m},
\end{eqnarray}
if $pq <1$ and $\phi_2(p,q)<\phi_1(p,q)$.
\begin{eqnarray}
\label{dinse}
a^{\dag m} a^n
=q^{-\nu\binom{m}2}\Big[{_q}\mathcal{S}_{a^\dag}^{-1}\mathcal{T}_{q,p^{-1}}^m
(\phi_2(p,q))\Big]\Big(\frac{\phi_1(p,q)}{\phi_2(p,q)}(p q)^{-\alpha N};(pq)^{\nu}\Big)_m a^{n-m},
\end{eqnarray}
if $pq >1$ and $\phi_1(p,q)<\phi_2(p,q)$.

$\bullet\,\,\,\,\,\,$ If $n=m$,
\begin{eqnarray}
\label{saa2}
a^n a^{\dag n} = 
p^{-\nu\binom{n}2}\mathcal{T}_{p^{-1},q}^n(\phi_1(p,q))\Big(\frac{\phi_2(p,q)}{\phi_1(p,q)}(p q)^{\alpha N+\nu};(pq)^{\nu}\Big)_n,
\end{eqnarray}
if $pq <1$ and $\phi_2(p,q)<\phi_1(p,q)$.
\begin{eqnarray}
\label{saa2}
a^n a^{\dag n} = 
q^{\nu\binom{n}2}\mathcal{T}_{q,p^{-1}}^n(\phi_2(p,q))\Big(\frac{\phi_1(p,q)}{\phi_2(p,q)}(p q)^{-\alpha N-\nu};(pq)^{-\nu}\Big)_n,
\end{eqnarray}
if $pq <1$ and $\phi_2(p,q)<\phi_1(p,q)$.
\begin{eqnarray}
a^{\dag n} a^n=
p^{\nu\binom{n}2}\Big[{_p}\mathcal{S}_{a^\dag}^{-1}\mathcal{T}_{p^{-1},q}(\phi_1(p,q))\Big]^n\Big(\frac{\phi_2(p,q)}{\phi_1(p,q)}(p q)^{\alpha N};(pq)^{-\nu}\Big)_n,
\end{eqnarray}
if $pq <1$ and $\phi_2(p,q)<\phi_1(p,q)$.
\begin{eqnarray}
a^{\dag n} a^n=
q^{-\nu\binom{n}2}\Big[{_q}\mathcal{S}_{a^\dag}^{-1}\mathcal{T}_{q,p^{-1}}(\phi_1(p,q))\Big]^n\Big(\frac{\phi_2(p,q)}{\phi_1(p,q)}(p q)^{-\alpha N};(pq)^{\nu}\Big)_n,
\end{eqnarray}
if $pq >1$ and $\phi_1(p,q)<\phi_2(p,q)$.
In the above expressions, $ _{(p,q)}\mathcal{S}_{a^\dag}$ is the translation operator  defined as follows
\begin{eqnarray}
\,\,_{(p,q)}\mathcal{S}_{a^\dag}={_p}\mathcal{S}_{a^\dag}\, {_q}\mathcal{S}_{a^\dag},\quad
_{p}\mathcal{S}_{a^\dag}^np^{\alpha N}=p^{\alpha N+n\nu},\quad {_q}\mathcal{S}_{a^\dag}^n q^{\alpha N}=q^{\alpha N+n\nu},\quad \forall\, n\in\mathbb{N}.
\end{eqnarray}
 The operator $\mathcal{T}_{p,q}(t)$
acts on the vacuum state $|0\rangle$ as follows
\begin{eqnarray}
\mathcal{T}_{p,q}(t) |0\rangle=\frac{tp^{\alpha \chi_0+\nu}}{p^\nu-q^\nu}|0\rangle,
\end{eqnarray}
where the
 product $(a;q)_l=(1-a)(1-aq)\ldots(1-aq^{l-1})$,  $l=1, 2,3\ldots$. 
This results reveal to be useful for  deducing  the commutators
between $a^n$ and $a^{\dag m}$ 

$\bullet\,\,\,$ For $n < m$,
\begin{eqnarray*}
[a^n,a^{\dag m}]_{q^\nu}
&=&
p^{-\nu\binom{n}2}(1-q^\nu
{_{(p,q)}}\mathcal{S}_{a^\dag}^m)\mathcal{T}_{p^{-1},q}^n(\phi_1(p,q)) \cr
&\times&\Big(\frac{\phi_2(p,q)}{\phi_1(p,q)}(p q)^{\alpha N+\nu};(pq)^{\nu}\Big)_n a^{\dag m-n},
\end{eqnarray*}
 if $pq <1$ and $\phi_2(p,q)<\phi_1(p,q)$ and 
\begin{eqnarray*}
[a^n,a^{\dag m}]_{q^\nu}
&= &q^{\nu\binom{n}2}(1-q^\nu\mathcal{S}_{a^\dag}^{-m})\mathcal{T}_{q,p^{-1}}^n(\phi_2(p,q))\cr
&\times& 
 \Big(\frac{\phi_1(p,q)}{\phi_2(p,q)}(p q)^{-\alpha N-\nu};(pq)^{-\nu}\Big)_n a^{\dag m-n},
\end{eqnarray*}
if $pq >1$ and $\phi_1(p,q)< \phi_2(p,q).$ 

$\bullet\,\,\,$  For $n > m$,
\begin{eqnarray}
[a^n,a^{\dag m}]_{q^\nu}&=&p^{\nu\binom{m}2}(
{_{(p,q)}}\mathcal{S}_{a^\dag}^n-q^\nu )
{_p}\mathcal{S}_{a^\dag}^{-1}\mathcal{T}_{p^{-1},q}^m(\phi_1(p,q))\cr
&\times&\Big(\frac{\phi_2(p,q)}{\phi_1(p,q)}(p q)^{\alpha N};(pq)^{-\nu}\Big)_m a^{n-m},
\end{eqnarray}
 if $pq <1$ and $\phi_2(p,q)<\phi_1(p,q)$ and  
\begin{eqnarray}
[a^n,a^{\dag m}]_{q^\nu} &=& q^{-\nu\binom{m}2}({_{(p,q)}}\mathcal{S}_{a^\dag}^n-q^\nu)
\Big[{_q}\mathcal{S}_{a^\dag}^{-1}\mathcal{T}_{q,p^{-1}}^m
(\phi_2(p,q))\Big]\cr
&\times&\Big(\frac{\phi_1(p,q)}{\phi_2(p,q)}(p q)^{-\alpha N};(pq)^{\nu}\Big)_m a^{n-m},
\end{eqnarray}
 if $pq >1$ and $\phi_1(p,q)< \phi_2(p,q).$  Noting that by similar computation we can derive the $p^{-\nu}$ commutators between   $a^n$ and $a^{\dag m}$.
\section{Coherent  states }
In this part we  investigate some class of  deformed states. These states are the eigenstates of the annihilation operator so called coherent states.
\\
{\it{\bf Proposition3}
The  CS associated with the algebra \eqref{sa7} and \eqref{eqref} with
  ($\chi_0=0$  and $\nu=\alpha$ ) are given in different case  by
\begin{enumerate}
\item if $pq<1,\;\phi_2(p,q)<\phi_1(p,q)$
\begin{eqnarray}
\label{shd}
|z\rangle_{\nu}=\mathcal{N}_{\nu}^{-1/2}(|z|^2)
\sum_{n=0}^\infty
\frac{p^{n\nu(n-1)/4}z^n}{\sqrt{\tau^n(\phi_1^{-1}(p,q)\phi_2(p,q)(pq)^{\nu};(pq)^\nu)_n}}|n\rangle,\,z\in\mathcal{D}_{\nu},
\end{eqnarray}
where
\begin{eqnarray}
\label{sehd}
\mathcal{N}_{\nu}(x)=
\sum_{n=0}^\infty
\frac{p^{\nu\binom{n}2}}{(\phi_1^{-1}(p,q)\phi_2(p,q)(pq)^{\nu};(pq)^\nu)_n}\Big(\frac{1-(pq)^{\nu}}{\phi_1(p,q)}x\Big)^n,
\end{eqnarray}
\item if $pq>1$, $\phi_1(p,q)<\phi_2(p,q)$
\begin{eqnarray}
\label{shde}
|z\rangle_{\nu}=\mathcal{N}_{\nu}^{-1/2}(|z|^2)
\sum_{n=0}^\infty
\frac{q^{-n\nu(n-1)/4}z^n}{\sqrt{\tau^n(\phi_2^{-1}(p,q)\phi_1(p,q)(pq)^{-\nu};(pq)^{-\nu})_n}}|n\rangle,\,z\in\mathcal{D}_{\nu},
\end{eqnarray}
where
\begin{eqnarray}
\mathcal{N}_{\nu}(x)=\sum_{n=0}^\infty
\frac{q^{-\nu\binom{n}2}}{(\phi_2^{-1}(p,q)\phi_1(p,q)(pq)^{-\nu};(pq)^{-\nu})_n}\Big(\frac{1-(pq)^{-\nu}}{\phi_2(p,q)}x\Big)^n,
\end{eqnarray}
\end{enumerate}
In the above expressions the convergence domain $\mathcal{D}_{\nu}$
of the serie $\mathcal{N}_{\nu}(x)$ is given by
\begin{eqnarray}
\label{seehd}
\mathcal{D}_{\nu}=
\{z\in\mathbb{C}:|z|^2<R_{\nu}\}, \,\mbox{ with } 
R_{\nu}=\infty,\;\nu>0.
\end{eqnarray}
$R_{\nu}$ is the radius associated with the same series.
}

In the limit,  when $\phi_1(p,q)=1=\phi_1(p,q)$, $p,\,q\rightarrow 1$, the series $\mathcal{N}_{\nu}(x)$ is reduced to the usual exponential function $e^{x}$. Also,
when $\phi_1(p,q)=1=\phi_1(p,q)$, $p \rightarrow 1,\nu \rightarrow 1$ the series $\mathcal{N}_{\nu}(x)$ is reduced to the $q-$exponential function 
$e_q(x)$. The most important of property of $|z\rangle_{\nu}$ is  that if $|z'\rangle_\nu$ is another  
CS, then
\begin{eqnarray}
\label{prod}
_\nu\langle z'|z\rangle_\nu
=\frac{\mathcal{N}_{\nu}(z\bar{z}')}{\sqrt{\mathcal{N}_{\nu}(|z|^2)\mathcal{N}_{\nu}(|z'|^2)}},
\end{eqnarray}
which means that such states  are not orthogonal.
\\
{\it {\bf Proposition}
The CS defined in \eqref{shd} and \eqref{shde}
are normalized,
are continuous in $z$. 
and solve the unity i.e, 
$
\int_{\mathcal{D}_{\nu}}\frac{d^2}{\pi}\mathcal{W}_\nu(|z|^2)|z\rangle_\nu \, {_\nu}\langle z|={\bf 1}.
$
The resolution of unity assumes the existence of a positive weight
a
function $\mathcal{W}_\nu(|z|^2)$ such that
$\mathcal{W}_\nu(x)=\mathcal{N}_\nu(x)
\tilde{\mathcal{W}}_\nu(x),\;x=|z|^2.$}

{\bf Proof:}  From \eqref{prod}, 
when $z'\to z, $ \eqref{shd} and \eqref{shde} are normalized. For the continuity we can see that
$
|||z'\rangle_{\nu}-|z\rangle_{\nu}||^2=2(1-\mathcal{R}e\,_\nu\langle z'|z\rangle_{\nu}),
$
so 
$$
|||z'\rangle_{\nu}-|z\rangle_{\nu}||^2\rightarrow 0 \mbox { as }  |z-z'|\rightarrow 0,
$$
since $_{\nu}\langle z'|z\rangle_{\nu}\rightarrow 0$ as  $|z'-z |\rightarrow 0.$
The resolution of the unity can be seem by using the relation
\begin{eqnarray*}
\int_{\mathcal{D}_{\nu}}\frac{d^2}{\pi}\mathcal{W}_\nu(|z|^2)|z\rangle_\nu \, {_\nu}\langle z|&=&
\sum_{n,m=0}^\infty
\frac{p^{n\nu(n-1)/4+m\nu(m-1)/4}\tau^{-\frac{n+m}{2}}}{\sqrt{(\frac{\phi_2(p,q)}{\phi_1(p,q)}(pq)^{\nu};(pq)^\nu)_n(\frac{\phi_2(p,q)}{\phi_1(p,q)}(pq)^{\nu};(pq)^\nu)_m}}\cr
&\times&\int_{\mathcal{D}_{\nu}}
\bar{z}^mz^n\frac{d^2\mathcal{W}_\nu(|z|^2)}{\pi \mathcal{N}_{\nu}(|z|^2)}.
\end{eqnarray*}
where $\tilde{\mathcal{W}}_\nu(x)$ has to be determined from the equations
\begin{eqnarray}
\int_{\mathcal{D}_{\nu}}x^n 
\tilde{\mathcal{W}}_\nu(x)dx=p^{-n\nu(n-1)/2}\Big(\frac{\phi_2(p,q)}{\phi_1(p,q)}(pq)^{\nu};(pq)^\nu\Big)_n\Big(\frac{\phi_1(p,q)}{1-(pq)^\nu}\Big)^n,
\end{eqnarray}
if $pq<1,\;\phi_2(p,q)<\phi_1(p,q)$ and 
\begin{eqnarray}
\int_{\mathcal{D}_{\nu}}x^n 
\tilde{\mathcal{W}}_\nu(x)dx =q^{n\nu(n-1)/2}\Big(\frac{\phi_1(p,q)}{\phi_2(p,q)}(pq)^{-\nu};(pq)^{-\nu}\Big)_n\Big(\frac{\phi_2(p,q)}{1-(pq)^{-\nu}}\Big)^n,
\end{eqnarray}
if $pq>1,\;\phi_1(p,q)<\phi_2(p,q)$. If $n$ is extended to $s-1,\;s\in\mathbb{C},$ then the problem
can be
formulated  to the classical Stieltjes power moment problem when $0< pq<1$ or Hausdorff power moment problem when 
$pq>1. \;\blacksquare$

\section{Matrix elements}
Let us now compute correlation functions with matrix elements of normal and antinormal forms  pertaining to quantum optics.\\
$\bullet$ For $n<m$, if $pq<1,\;\phi_2(p,q)<\phi_1(p,q),$
the normal form 
 is defined by
\begin{eqnarray}
_\nu\langle z|a^{\dagger m}a^n|z\rangle_\nu
= \mathcal{N}_{\nu}^{-1}(|z|^2)\sum_{r,s=0}^\infty\frac{p^{r\nu(r-1)/4+s\nu(s-1)/4}\tau^{-\frac{r+s}{2}}}{\sqrt{(\frac{\phi_2(p,q)}{\phi_1(p,q)}(pq)^{\nu};(pq)^\nu)_r(\frac{\phi_2(p,q)}{\phi_1(p,q)}(pq)^{\nu};(pq)^\nu)_s}}\mathcal{F}_{mn}^{rs},
\end{eqnarray}
where  the matrix elements $\mathcal{F}_{mn}^{rs}$ are given by
\begin{eqnarray}
\label{vf}
&\mathcal{F}_{mn}^{rs}=\langle r| a^{\dag m} a^n |s\rangle\nonumber\\
&=C_sC_{m-n+s}^{-1}p^{\nu\binom{n}2}\Big(\frac{\phi_1(p,q)p^{(1-s)\nu}}{1-(pq)^\nu}\Big)^n\Big(\frac{\phi_2(p,q)}{\phi_1(p,q)}(pq)^{s\nu};(pq)^{-\nu}\Big)_n\delta_{r,m-n+s},
\end{eqnarray}
and
\begin{eqnarray}
C_n^2=p^{\nu\binom{n}2}\Big(\frac{1-(pq)^\nu}{\phi_1(p,q)}\Big)^n\Big(\frac{\phi_2(p,q)}{\phi_1(p,q)}(pq)^{\nu};(pq)^\nu\Big)_n^{-1},
\end{eqnarray}
if $pq>1,\;\phi_1(p,q)< \phi_2(p,q)$
\begin{eqnarray}
_\nu\langle z|a^{\dagger m}a^n|z\rangle_\nu
= \mathcal{N}_{\nu}^{-1}(|z|^2)\sum_{r,s=0}^\infty\frac{q^{-r\nu(r-1)/4-s\nu(s-1)/4}\tilde{ \tau}^{-\frac{r+s}{2}}\quad \mathbf{F}_{mn}^{rs}}{\sqrt{(\frac{\phi_1(p,q)}{\phi_2(p,q)}(pq)^{-\nu};(pq)^{-\nu})_r(\frac{\phi_1(p,q)}{\phi_2(p,q)}(pq)^{-\nu};(pq)^{-\nu})_s}},\nonumber
\end{eqnarray}
where the matrix elements $\mathbf{F}_{mn}^{rs}$ are given by
\begin{eqnarray}
\label{vf}
&\mathbf{F}_{mn}^{rs}=\langle r| a^{\dag m} a^n |s\rangle\nonumber\\
&=\tilde{C}_s\tilde{C}_{m-n+s}^{-1}q^{-\nu\binom{n}2}\Big(\frac{\phi_2(p,q)q^{(s-1)\nu}}{1-(pq)^{-\nu}}\Big)^n\Big(\frac{\phi_1(p,q)}{\phi_2(p,q)}(pq)^{-s\nu};(pq)^{\nu}\Big)_n\delta_{r,m-n+s},
\end{eqnarray}
\begin{eqnarray}
\tilde{C}_n^2=q^{-\nu\binom{n}2}\Big(\frac{1-(pq)^{-\nu}}{\phi_2(p,q)}\Big)^n\Big(\frac{\phi_1(p,q)}{\phi_2(p,q)}(pq)^{-\nu};(pq)^{-\nu}\Big)_n^{-1}.
\end{eqnarray}
The antinormal form if $pq<1,\;\phi_2(p,q)<\phi_1(p,q),$
is given by
\begin{eqnarray}
_\nu\langle z|a^na^{\dag m}|z\rangle_\nu
= \mathcal{N}_{\nu}^{-1}(|z|^2)\sum_{r,s=0}^\infty\frac{p^{r\nu(r-1)/4+s\nu(s-1)/4}\tau^{-\frac{r+s}{2}}}{\sqrt{(\frac{\phi_2(p,q)}{\phi_1(p,q)}(pq)^{\nu};(pq)^\nu)_r(\frac{\phi_2(p,q)}{\phi_1(p,q)}(pq)^{\nu};(pq)^\nu)_s}}\mathcal{G}_{mn}^{rs}
\end{eqnarray}
where the matrix elements $\mathcal{G}_{mn}^{rs}$ are given by
\begin{eqnarray}
\label{dg}
&\mathcal{G}_{mn}^{rs}=\langle r| a^n a^{\dag m}  |s\rangle\nonumber\\
&=C_sC_{m-n+s}^{-1}p^{\nu\binom{n}2}\Big(\frac{\phi_1(p,q)p^{(1-s-m)\nu}}{1-(pq)^\nu}\Big)^n\Big(\frac{\phi_2(p,q)}{\phi_1(p,q)}(pq)^{(s+m)\nu};(pq)^{-\nu}\Big)_n\delta_{r,m-n+s}.
\end{eqnarray}
if $pq>1,\;\phi_1(p,q)<\phi_2(p,q)$
\begin{eqnarray*}
_\nu\langle z|a^na^{\dag m}|z\rangle_\nu
= \mathcal{N}_{\nu}^{-1}(|z|^2)\sum_{r,s=0}^\infty\frac{q^{-r\nu(r-1)/4-s\nu(s-1)/4}\tilde{\tau}^{-\frac{r+s}{2}}\;\;\mathbf{G}_{mn}^{rs}}{\sqrt{(\frac{\phi_1(p,q)}{\phi_2(p,q)}(pq)^{-\nu};(pq)^{-\nu})_r(\frac{\phi_1(p,q)}{\phi_2(p,q)}(pq)^{-\nu};(pq)^{-\nu})_s}}
\end{eqnarray*}
where   the matrix elements $\mathbf{G}_{mn}^{rs}$ are given by
\begin{eqnarray}
\label{dg}
&\mathbf{G}_{mn}^{rs}=\langle r| a^n a^{\dag m}  |s\rangle\nonumber\\
&=\tilde{C}_s\tilde{C}_{m-n+s}^{-1}q^{-\nu\binom{n}2}\Big(\frac{\phi_2(p,q)q^{(s+m-1)\nu}}{1-(pq)^{-\nu}}\Big)^n\Big(\frac{\phi_1(p,q)}{\phi_2(p,q)}(pq)^{-(s+m)\nu};(pq)^{\nu}\Big)_n\delta_{r,m-n+s}.
\end{eqnarray}

$\bullet$ For $n>m,$ if $pq<1,\;\phi_2(p,q)<\phi_1(p,q),$
the normal form  is defined by
\begin{eqnarray}
_\nu\langle z|a^{\dag m}a^n|z\rangle_\nu= \mathcal{N}_{\nu}^{-1}(|z|^2)\sum_{r,s=0}^\infty\frac{p^{r\nu(r-1)/4+s\nu(s-1)/4}\tau^{-\frac{r+s}{2}}}{\sqrt{(\frac{\phi_2(p,q)}{\phi_1(p,q)}(pq)^{\nu};(pq)^\nu)_r(\frac{\phi_2(p,q)}{\phi_1(p,q)}(pq)^{\nu};(pq)^\nu)_s}}\mathcal{\tilde{G}}_{mn}^{rs},
\end{eqnarray}
where the matrix elements  $\mathcal{\tilde{G}}_{mn}^{rs}$ are given by
\begin{eqnarray}
&\mathcal{\tilde{G}}_{mn}^{rs}=\langle r| a^{\dag m}a^n |s\rangle\nonumber\\
&=C_rC_{r+n-m}^{-1}p^{-\nu\binom{m}2}\Big(\frac{\phi_1(p,q)p^{(n-s)\nu}}{1-(pq)^\nu}\Big)^n\Big(\frac{\phi_2(p,q)}{\phi_1(p,q)}(pq)^{(s-m+1)\nu};(pq)^{\nu}\Big)_m\delta_{s,r+n-m},
\end{eqnarray}
if $pq>1,\;\phi_1(p,q)<\phi_2(p,q),$
\begin{eqnarray}
_\nu\langle z|a^{\dag m}a^n|z\rangle_\nu
= \mathcal{N}_{\nu}^{-1}(|z|^2)\sum_{r,s=0}^\infty\frac{q^{-r\nu(r-1)/4-s\nu(s-1)/4}\tilde{ \tau}^{-\frac{r+s}{2}}\;\mathbf{\tilde{G}}_{mn}^{rs}}{\sqrt{(\frac{\phi_1(p,q)}{\phi_2(p,q)}(pq)^{-\nu};(pq)^{-\nu})_r(\frac{\phi_1(p,q)}{\phi_2(p,q)}(pq)^{-\nu};(pq)^{-\nu})_s}},\nonumber
\end{eqnarray}
where  the matrix elements $\mathbf{\tilde{G}}_{mn}^{rs}$ are given by
\begin{eqnarray}
&\mathbf{\tilde{G}}_{mn}^{rs}=\langle r|a^{\dag m}a^n|s\rangle\nonumber\\
&=\tilde{C}_r\tilde{C}_{r+n-m}^{-1}q^{\nu\binom{m}2}\Big(\frac{\phi_2(p,q)q^{(s-n)\nu}}{1-(pq)^{-\nu}}\Big)^n\Big(\frac{\phi_1(p,q)}{\phi_2(p,q)}(pq)^{(n-s-1)\nu};(pq)^{-\nu}\Big)_m\delta_{s,r+n-m},
\end{eqnarray}
The  antinormal form if $pq<1,\;\phi_2(p,q)<\phi_1(p,q),$
is expressed by the formula:
\begin{eqnarray}
_\nu\langle z|a^na^{\dag m}|z\rangle_\nu
= \mathcal{N}_{\nu}^{-1}(|z|^2)\sum_{r,s=0}^\infty\frac{p^{r\nu(r-1)/4+s\nu(s-1)/4}\tau^{-\frac{r+s}{2}}}{\sqrt{(\frac{\phi_2(p,q)}{\phi_1(p,q)}(pq)^{\nu};(pq)^\nu)_r(\frac{\phi_2(p,q)}{\phi_1(p,q)}(pq)^{\nu};(pq)^\nu)_s}}\mathcal{\tilde{F}}_{mn}^{rs},
\end{eqnarray}
where   the matrix elements $\mathcal{\tilde{F}}_{mn}^{rs}$ are given by
\begin{eqnarray}
\label{ftil}
&\mathcal{\tilde{F}}_{mn}^{rs}=\langle r|a^na^{\dag m}|s\rangle\nonumber\\
&=C_sC_{r+n-m}^{-1}p^{-\nu\binom{m}2}\Big(\frac{\phi_1(p,q)p^{-s\nu}}{1-(pq)^\nu}\Big)^n\Big(\frac{\phi_2(p,q)}{\phi_1(p,q)}(pq)^{(s+1)\nu};(pq)^{\nu}\Big)_m\delta_{s,r+n-m}.
\end{eqnarray}
if $pq>1,\;\phi_1(p,q)<\phi_2(p,q)$
\begin{eqnarray}
_\nu\langle z|a^n a^{\dag m}|z\rangle_\nu
= \mathcal{N}_{\nu}^{-1}(|z|^2)\sum_{r,s=0}^\infty\frac{q^{-r\nu(r-1)/4-s\nu(s-1)/4}\tilde{ \tau}^{-\frac{r+s}{2}}\;\mathbf{\tilde{G}}_{mn}^{rs}}{\sqrt{(\frac{\phi_1(p,q)}{\phi_2(p,q)}(pq)^{-\nu};(pq)^{-\nu})_r(\frac{\phi_1(p,q)}{\phi_2(p,q)}(pq)^{-\nu};(pq)^{-\nu})_s}},\nonumber
\end{eqnarray}
where  the matrix elements $\mathbf{\tilde{F}}_{mn}^{rs}$  are given by
\begin{eqnarray}
&\mathbf{\tilde{F}}_{mn}^{rs}=\langle r|a^n a^{\dag m}|s\rangle\nonumber\\
&=\tilde{C}_s\tilde{C}_{r+n-m}^{-1}q^{\nu\binom{m}2}\Big(\frac{\phi_2(p,q)q^{s\nu}}{1-(pq)^{-\nu}}\Big)^n\Big(\frac{\phi_1(p,q)}{\phi_2(p,q)}(pq)^{-(s+1)\nu};(pq)^{-\nu}\Big)_m\delta_{s,r+n-m}.
\end{eqnarray}
\section{Discussions}
This part addressed the study of the previous deformation in the particular case that  we present. 
Let us reexpressed the relations \eqref{sa7} as
\begin{eqnarray}
\label{sa15}
aa^{\dag n}-q^{n\nu}a^{\dag n}a &=&
[n\nu ]_{p,q,\nu}^{1,1}a^{\dag n-1}p^{-\alpha N}\cr
a a^{\dag n}-p^{-n\nu}a^{\dag n} a
&=& [n\nu]_{p,q,\nu}^{1,1}a^{\dag n-1}q^{\alpha N},\;\;\forall
 \,n\,\in\mathbb{N}\diagdown \{0\},
\end{eqnarray}
where the quantity $[n]_{p,q,\nu}^{\phi_1,\phi_2}$ is the deformed number    given by
\begin{eqnarray}
\label{asa}
[n\nu]_{p,q,\nu}^{\phi_1,\phi_2}=\frac{\phi_1(p,q)p^{-n\nu}-\phi_2(p,q)q^{n\nu}}{p^{-\nu}-q^\nu}.
\end{eqnarray}
 Then it appears important to emphasize that: By setting $C_1$ as the Casimir operator and  defined the regular operators   functions
\begin{eqnarray}
 \hat{\phi}_1(p,q,C_1)=(1+2\gamma C_1)p^{-\beta},\,\,\,\,\hat{\phi}_2(p,q,C_1)=(1+2\gamma C_1)q^{\beta}
\end{eqnarray}
such that $\hat{\phi}_1|n\rangle=(1+2\gamma \omega^2)p^{-\beta}|n\rangle$ and  $\hat{\phi}_2|n\rangle=(1+2\gamma \omega^2)q^{\beta}|n\rangle$, where $w^2$ is the eigenvalue of the operator $C_1$.
 The commutation relations \eqref{sa7} take the form
\begin{eqnarray}
\label{sa444}
aa^\dagger - q^\nu a^\dagger a =\hat{\phi}_1(p,q,C_1)
p^{-\alpha N },\,\,
 aa^\dagger - p^{-\nu} a^\dagger a = \hat{\phi}_2(p,q,C_1)
q^{\alpha N }.
\end{eqnarray}
The eigen-equation associated with the 
Hamiltonian operator $H_{\alpha,\beta}^{\nu, \gamma} (p, q) =
a^\dagger a + aa^\dagger,$  i.e.
$
\label{aj}
H_{\alpha,\beta}^{\nu, \gamma} (p, q)|n\rangle=
E_{n,\alpha,\beta}^{\nu,\gamma} (p, q)|n\rangle
$, is such that the corresponding  eigenvalue  $E_{n,\alpha,\beta}^{\nu,\gamma} (p, q)$ is written as
\begin{eqnarray}
\label{ea}
E_{n,\alpha,\beta}^{\nu,\gamma} (p, q)&=&
(1+2\gamma w^2)\{q^{\alpha\chi_0+n\nu+\beta}+
(1+p^{-\nu})[ n\nu+\alpha\chi_0+\beta]_{(p,q;\nu)}\}
\end{eqnarray}
or, equivalently,
\begin{eqnarray}\label{eqeq}
E_{n,\alpha,\beta}^{\nu,\gamma} (p, q)&=&(1+2\gamma w^2)\{p^{-\alpha\chi_0 -n\nu-\beta}+
(1+q^{\nu})[ n\nu+\alpha\chi_0+\beta]_{(p,q;\nu)}\},
\end{eqnarray}
where
\begin{eqnarray}
[n;\gamma C_1]_{(p,q;\nu)}!=\Big(1+2\gamma C_1\Big)^n\frac{((p^{-\nu},q^\nu);(p^{-\nu},q^\nu))_n}{(p^{-\nu}-q^\nu)^n}.
\end{eqnarray}
 Indeed,
for $n, m\in\mathbb{N}\diagdown\{0\}$, we  derive
the next commutators: 

$\bullet\,\,\,$ For $n < m,\,p<1$ and $q>1$, 
\begin{eqnarray*}
[a^n,(a^\dagger)^m]_{q^\nu} 
&=&\frac{ (1+2\gamma C_1)^n}{(p^{-\nu}-q^\nu)^n}\Bigg[
((p^{-\alpha N-\beta-\nu},
q^{\alpha N+\beta+\nu});(p^{-\nu},
q^\nu))_n\cr
&-&q^\nu((p^{-\alpha N-\beta+(m-1)\nu},
q^{\alpha N+\beta-(m-1)\nu});(p^{-\nu},
q^\nu))_n\Bigg](a^\dagger)^{m-n}.
\end{eqnarray*}
There follows:
\begin{align}
\label{samadine}
\sum\limits_{n=0}^{m-1}\frac{[a^n,a^{\dagger m}]_{q^\nu}}
{[n;\gamma C_1]_{(p,q;\nu)}!}={_{(p,q)}}\mathcal{S}_{a^\dagger}\Big(1-q^\nu _{(p,q)}\mathcal{S}_{a^\dagger}^ {-m}\Big)\sum\limits_{n=0}^{m-1}\frac{((p^{-\alpha N-\beta},
q^{\alpha N+\beta});(p^{-\nu},
q^\nu))_n}{((p^{-\nu},q^\nu);(p^{-\nu},q^\nu))_n}{a^\dagger}^{m-n}\cr
={_{(p,q)}}\mathcal{S}_{a^\dagger}\Big(1-q^\nu _{(p,q)}\mathcal{S}_{a^\dagger}^ {-m}\Big)\mathcal{L}_{m-1}
[(p^{-\alpha N-\beta},q^{\alpha N+\beta});(p^{-\nu},q^\nu);{a^\dagger}^{-1}]{a^\dagger}^{m},
\end{align}
where $\mathcal{L}_{m}$ is a deformed hypergeometric function given by
\begin{eqnarray}
\mathcal{L}_{m}[(\lambda,\sigma);(p,q);z]:=\sum\limits_{n=0}^m\frac{((\lambda,\sigma);(p,q))_n}{((p,q);(p,q))_n}z^n,
\end{eqnarray}
By using the $(p,q)$-binomial theorem given by
\begin{eqnarray}
\label{mn}
\sum\limits_{n=0}^{\infty}\frac{((a,b);(p,q))_n}{((p,q);(p,q))_n}z^n=\frac{((p,bz);(p,q))_\infty}{((p,az);(p,q))_\infty},
\end{eqnarray}
and for $m\rightarrow \infty,$ the relation \eqref{samadine} is reduced to the  
\begin{eqnarray}
\mathcal{L}_{\infty}
&=&\frac{\Big((p^{-\nu}, q^{\alpha N+\beta}
a^{\dagger-1});(p^{-\nu},q^{\nu})\Big)_\infty}{((p^{-\nu}, p^{-\alpha N-\beta}
a^{\dagger-1});(p^{-\nu},q^{\nu}))_\infty},\quad ||a^{\dagger-1}|| <1,\cr
&&
\end{eqnarray}
$\bullet\,\,\,$  For $n > m$,
by using  the identity
\begin{eqnarray}
 [m;\gamma C_1]_{(p,q;\nu)}!=
(-1)^m(1+2\gamma C_1)^m(p^{-1}q)^{m\nu+\nu m(m-1)/2}\frac{((p^\nu,
q^{-\nu});(p^\nu,q^{-\nu}))_m}{(p^{-\nu}-q^\nu)^m},
\end{eqnarray}
 we infer
\begin{eqnarray}
\sum\limits_{m=0}^{n-1}\frac{[a^n,a^{\dagger m}]_{q^\nu}
((p^\nu,0);(p^\nu,q^{-\nu}))_m}{[m;\gamma C_1]_{(p,q;\nu)}!}
=(_{(p,q)}T_a^n-q^\nu)\mathcal{L}_{n-1}[(p^{-\alpha N-\beta},q^{\alpha N+\beta});(p^\nu,q^{-\nu});a^{-1}]a^n,\cr
\end{eqnarray}
where
\begin{eqnarray}
\label{right}
\mathcal{L}_{n-1}[(p^{-\alpha N-\beta},q^{\alpha N+\beta});(p^\nu,q^{-\nu});a^{-1}]=\sum\limits_{m=0}^{n-1}\frac{((p^{-\alpha N-\beta},q^{\alpha N+\beta});(p^\nu,q^{-\nu}))_m}{((0,q^{-\nu}),(p^\nu,q^{-\nu});(p^\nu,q^{-\nu}))_m}a^{-m}.
\end{eqnarray}
When $n\rightarrow \infty$, \eqref{right} becomes
\begin{eqnarray}
\mathcal{L}_{\infty}&=&{_2}\phi_1\left(\left.\begin{array}{c} (p^{-\alpha N-\beta},q^{\alpha N+\beta}),0\\
(0,q^{-\nu}) \end{array}\right\vert(p^{\nu},
q^{-\nu});a^{-1}\right) ||a^{-1}||<1.\cr
&&
\end{eqnarray}
$\bullet\,\,\,$  For $n=m,$ 
 we obtain
 $$ \sum\limits_{n=0}^\infty {{((1,0),(p^{-\nu},q^\nu))_n
a^n(a^\dagger)^{2n}a^n(p^{-\nu}-q^\nu)^{n}} \over
{[n;\gamma C_1]!}}\; =$$
\begin{eqnarray}
\label{aa2s}
{_2}\phi_1\Bigg(\left.\begin{array}{cc}
(p^{-\alpha N-\beta-\nu},q^{\alpha N+\beta+\nu}),
(p^{\alpha N+\beta},q^{-\alpha N-\beta})\\
(0,1)\end{array}\right\vert(p^{-\nu},q^\nu);(1+2\gamma C_1)(qp^{-1})^{\alpha N+\beta}\Bigg),
\end{eqnarray}
\section{Conclusion}
In this paper, associated relevant properties  have been investigated for the induced deformed harmonic oscillator; the energy spectrum   has been explicitly
 computed. Deformed coherent states have been built and discussed
with respect to the  criteria of coherent state construction.  Various commutators involving annihilation and creation operators and 
their combinatorics have been computed and analyzed. Finally, the correlation functions of matrix elements of main normal and antinormal forms, pertinent for  quantum optics
 analysis,  have been computed.

\section*{ Acknowledgments}
This work is partially supported by the Abdus Salam International
Centre for Theoretical Physics (ICTP, Trieste, Italy) through the
Office of External Activities (OEA) - \mbox{Prj-15}. The ICMPA
is also in partnership with
the Daniel Iagolnitzer Foundation (DIF), France. The authors thank Plyushchay M. S. for the fruitful discussions. D. Ousmane Samary thanks the Centre
international de math\'ematiques pures et appliqu\'ees (CIMPA) for financial supports.

\end{document}